%% file: ZeroInflatedDPMix_Oganisian.tex
\documentclass[10pt]{article}

\usepackage[left=24mm,top=1in,bottom=1in,right=24mm]{geometry}
\usepackage{setspace}
\usepackage{graphicx}
\usepackage{subfig}
\usepackage{amssymb}
\usepackage{amsmath}
\usepackage{hyperref}
\usepackage[font=small,skip=0pt]{caption}
\usepackage{mathrsfs}
\usepackage{bm}
\usepackage{titling}
\usepackage{lipsum}
\usepackage{algorithm}
\usepackage{caption}
\usepackage[noend]{algpseudocode}
\usepackage[english]{babel}
\usepackage{authblk}
\usepackage{tablefootnote}
\usepackage{booktabs}
\usepackage{calc}
\usepackage{makecell}
\usepackage{multirow}

\pretitle{\begin{center}\Large}
\posttitle{\par\end{center}\vskip 0.5em}
\preauthor{\begin{center}}
\postauthor{\end{center}}
\predate{\par\large\centering}
\postdate{\par}


\newlength\myindent
\numberwithin{equation}{section}

\title{\bf A Bayesian Nonparametric Model for Zero-Inflated Outcomes: Prediction, Clustering, and Causal Estimation}

\author[1]{Arman Oganisian \thanks{Corresponding Author. Email:\href{aoganisi@upenn.edu}{aoganisi@upenn.edu}.} }
\author[1]{Nandita Mitra}
\author[2]{Jason A. Roy }
\affil[1]{University of Pennsylvania \protect\\ Division of Biostatistics \protect\\ Department of Biostatistics, Epidemiology, and Informatics \vspace*{.5cm}}
\affil[2]{Rutgers University \protect\\ Department of Biostatistics and Epidemiology}

\date{\today}
\begin{document}
\onehalfspacing

\maketitle
\thispagestyle{empty}

\begin{abstract}
Researchers are often interested in predicting outcomes, conducting clustering analysis to detect distinct subgroups of their data, or computing causal treatment effects. Pathological data distributions that exhibit skewness and zero-inflation complicate these tasks - requiring highly flexible, data-adaptive modeling. In this paper, we present a fully nonparametric Bayesian generative model for continuous, zero-inflated outcomes that simultaneously predicts structural zeros, captures skewness, and clusters patients with similar joint data distributions. The flexibility of our approach yields predictions that capture the joint data distribution better than commonly used zero-inflated methods. Moreover, we demonstrate that our model can be coherently incorporated into a standardization procedure for computing causal effect estimates that are robust to such data pathologies. Uncertainty at all levels of this model flow through to the causal effect estimates of interest - allowing easy point estimation, interval estimation, and posterior predictive checks verifying positivity, a required causal identification assumption. Our simulation results show point estimates to have low bias and interval estimates to have close to nominal coverage under complicated data settings. Under simpler settings, these results hold while incurring lower efficiency loss than comparator methods. Lastly, we use our proposed method to analyze zero-inflated inpatient medical costs among endometrial cancer patients receiving either chemotherapy and radiation therapy in the SEER medicare database.
\end{abstract}

\newpage
\section{Introduction}
Researchers in many fields are often interested in tasks such as outcome prediction, clustering analysis, and causal inference. For example, researchers in personalized medicine are broadly concerned with forming out-of-sample outcome predictions given a new subject's covariates. Health economists are often interested in identification and analysis of sub-clusters of their data and may turn to algorithms such as K-means\cite{Liao2016ClusterAA}. Policy researchers, on the other hand, are particularly interested in causality - estimating the average difference in outcomes that would have occurred under hypothetical policy interventions. All of these tasks are complicated in the presence of zero-inflated outcomes, multi-modality, and extreme skewness. Structural zeros often need to be modeled - especially if the goal is causal effect estimation of a treatment that may itself influence the prevalence of zeros. For prediction purposes, it is also necessary to both capture outcomes at the skewed high-end of the distribution as well as predicting the structural zeros at the low-end. For robust clustering, we would prefer a method that does not require us to pre-specify the number of clusters in our data - which is often unknown to researchers.  \\

In this paper, we develop of a Bayesian nonparametric (BNP) generative model that simultaneously predicts structural zeros as a function of covariates, captures skewness in both the outcome and continuous covariates, and induces a grouping of subjects into (infinitely many) clusters with similar joint data distributions. The result is a flexible, multi-purpose model that is broadly applicable to the tasks described above. We demonstrate the utility of our approach for causal effect estimation in these difficult settings by developing a standardization procedure around the proposed model. This fully Bayesian approach allows uncertainty to propagate through to the causal estimates - allowing point and interval estimation of various causal contrasts such as mean differences and quantile causal effects. Moreover, posterior predictive checks around positivity - a key causal identification assumption - can be readily conducting using the model output. \\

In particular, we propose a Dirichlet Process (DP) mixture of zero-inflated regressions. Each zero-inflated regression is a two-part model: a logistic model for the probability of the outcome being zero and a Gaussian regression for the continuous, non-zero outcomes. DP mixtures \cite{ferguson1973, ferguson1983} are a class of BNP models that partition a complex joint distribution of the outcome and covariates into more homogeneous clusters. In our case, the cluster-specific conditional means are modeled using a two-part zero-inflated regression. Unlike finite mixtures, DP mixtures assume there are infinitely many clusters - removing the need to specify the number of clusters in advance. As many clusters are introduced as are needed to accommodate the complexity of the data. If the data are not complex and adequately fit using a parameteric model, new clusters form less often and our model shrinks to a simple parametric form. In this sense, our model is data adaptive - growing in proportion to the complexity of the data. \\

The flexibility and relative ease of constructing point and interval estimates for various types of contrasts are perhaps some of the reasons that BNP methods have been growing in popularity within causal inference. For example, Bayesian additive regression trees (BART)\cite{chipman2010, HIll2011} have been used to estimate average treatment effects. Dependent Dirichlet process (DDP) methods have been developed for estimating marginal structural models \cite{Roy2016}. Dirichlet process (DP) mixture approaches for mediation analysis\cite{Kim2017} and Enriched Dirichlet process (EDP) \cite{Wade2014} mixture approaches to standardization have also been developed \cite{Roy2018}. However, these methods are not suited for computing causal effects on zero-inflated outcomes. This paper contributes to the existing literature by developing a BNP standardization approach that accounts for zero-inflation. \\ 

At the same time, several factors distinguish our approach from the existing zero-inflated models outside of the causal inference literature. As opposed to the parametric Bayesian approach of Ghosh et al \cite{ghosh2006}, our method is non-parametric and, therefore, better suited for complex data. Barcella et al \cite{barcella2016} develop a DP mixture of poisson regressions. Though this provides a flexible fit to count data, it is inappropriate for semi-continuous data. The still unpublished work of Linero et al \cite{Linero2018} develops a semi-parametric Bayesian model for semi-continuous outcomes. They use a two-part model - a probit model for the probability of a zero and a parametric density for non-zero outcomes. The mean functions of both models are jointly estimated using a BART-based model. In contrast, our model is fully nonparametric, DP-based as opposed to BART-based, and generative as opposed to conditional. The strength of DP-based procedures over BART-based procedures is that the former induces clustering - allowing us to capture multi-modalities. Using generative models as opposed to conditional models provides a framework for flexibly imputing missing data - as was demonstrated by Roy et al 2018 \cite{Roy2018}.\\

Though broadly applicable, we motivate our approach throughout the paper by the analysis of medical cost outcomes - an important use case of our method. Zero-inflation is the norm in cost data as patients may tend to have zero costs in ways that depend on measured covariates and the assigned treatment. Medical costs also tend to be skewed by especially high-cost patients. Moreover, the joint distribution tends to be multi-modal with groups of patients that exhibit different cost-covariate relationships. Policymakers such as legislators and regulators such as the Food and Drug Administration often make use of economic analyses comparing costs between existing and proposed treatments. These comparisons are causal in nature and require robust statistical modeling while adjusting for confounders. \\

In the next section, we present the details of our generative model - including its posterior, hyperparameters, and an MCMC algorithm for posterior sampling. In Section 3, we outline a standardization procedure for computing causal effects using the posterior predictive distribution. We describe a Monte Carlo method that carries out standardization by ensembling predictions from the cluster-specific models and averaging them over the confounder distribution. We end the section by describing posterior predictive checks for positivity violations - an important check when making causal inference. \\

Section 4 presents simulation studies exploring bias, coverage, and precision of causal effect estimates from the zero-inflated DP model. We compare our model to alternative approaches that researchers may consider when encountered with zero-inflated cost data: BART, doubly robust estimators, and Gamma hurdle models. We show that under complex data distribution our model yields low bias posterior mean estimates of causal effects with close to nominal coverage. In simpler data distributions, our model also performs well with lower efficiency loss than comparator methods.  \\

In section 6, we apply our method to inpatient medical cost outcomes among patients with endometrial cancer identified in the SEER Medicare database. We show that our model is able to capture real cost distributions better than comparator methods by both predicting structural zeros as well as capturing skewness and multi-modality. The model's induced clustering detects unique subgroups of patients with different cost profiles. We present these results and end with computing various causal estimates contrasting inpatient costs among patients who were assigned to either chemotherapy or radiation therapy at diagnosis.

\section{Dirichlet Process Mixture of Zero-Inflated Regressions}
\subsection{A Generative Model}
\label{sc:genmod}
Consider observing data $ D = (Y_i, A_i, L_i)_{i=1:n}$ from $n$ independently sampled subjects. The $p \times 1$ covariate vector $L_i$ contains a mix of categorical and continuous covariates measured pre-treatment. The scalar $A_i\in \{0,1\}$ denotes binary treatment assignment. The scalar outcome is $Y_i$ - which is zero-inflated and may be skewed and multimodal. We also define an indicator $Z_i=1$ if $Y_i$ is zero and $Z_i=0$ otherwise.\\

In order to propose our generative model, we first define regression model vectors for subject $i$ as $x_i = (1, A_i, L_i)'$ and $m_i = (1, L_i)'$. We specify our model over the joint data distribution hierarchically as
\begin{equation}
\begin{split}\label{eq:genmod}
Y_i | A_i, L_i, \beta_i, \gamma_i, \phi_i & \sim \pi( x_i'\gamma_i )\delta_0(y_i) + \big(1 - \pi( x_i'\gamma_i ) \big) \cdot N(y_i| x_i'\beta_i, \phi_i   )  \\
A_i | L_i  & \sim Ber\Big( \text{expit}(\ m_i'\eta_i) \ \Big ) \\
L_i | \theta_i & \sim p(l_i | \theta_i) \\
\omega_i | G & \sim G \\
G & \sim DP(\alpha G_0)
\end{split}
\end{equation}

Above, we define $\omega_i =(\beta_i,\phi_i, \gamma_i, \eta_i, \theta_i)$ for compactness. The conditional distribution of the outcome, $Y_i$, is modeled as a two-part mixture of a point-mass at 0, $\delta_0(y_i) = I(y_i=0)$, and a Gaussian distribution with mean $x_i' \beta_i$ and variance $\phi_i$. This induces a positive probability on the outcome being zero, $P(Y_i=0) = \pi(x_i'\gamma_i) = expit(x_i'\gamma_i)$. This probability is modeled as a function of treatment and confounders using a logistic regression with a $(p+2) \times 1$ parameter vector $\gamma_i$. Separately, the conditional mean of non-zero outcomes is modeled using a regression with a $(p+2) \times 1$ parameter vector $\beta_i$.  \\

In anticipation of subsequent application to causal estimation, we explicitly model treatment probability (i.e., the propensity score) as a function of confounders, $L_i$, using a logistic regression with a $(p+1) \times 1$ parameter vector $\eta_i$. Finally, a joint distribution over the confounders, $p(l_i| \theta_i)$, is specified and governed by a vector of parameters $\theta_i$.\\

We assume subject-specific parameters are drawn from some distribution $G$. We place a DP prior with base distribution $G_0$ and concentration parameter $\alpha$ on $G$ - denoted as $G \sim DP(\alpha G_0)$. The DP is a ``distribution over distributions'' \cite{ferguson1973} and draws from a DP are discrete. The discreteness of $G$ implies a positive probability of ties among the subject-specific parameters. In other words, the DP prior induces a clustering of patients who are more homogeneous in terms of the parameters that govern the joint distribution of their data - including the conditional outcome distribution, the structural zero distribution, propensity score distribution, and covariate distribution. \\

The DP model assumes there are infinitely many clusters and, therefore, parameters - making this a nonparametric model despite the local parametric assumptions above \cite{hannah2011}. The model is data-adaptive in the sense that more clusters can arise to accommodate more complex data distributions. However, if the data are simple enough to be explained by a single set of parameters, then the model will likely reduce to the locally parametric one given above. \\

This model has many desirable properties when it comes to cost outcomes - our motivating use case for the method throughout the paper. The clustering accounts for multi-modality in cost distributions by partitioning patients into groups with similar joint distribution parameters. It also accounts for zero-inflation which is prevalent in cost data by explicitly modeling it as a function of treatment and confounders. In terms of causal inference, this model admits a flexible predictive distributions which, as we will see, can be incorporated into standardization procedure. Finally, explicit modeling of treatment assignment allows us to conduct posterior predictive checks assessing the validity of the positivity assumption.

\subsection{Posterior Sampling of Parameters}
\label{sc:postsample}

Using the chain rule, the joint prior on the parameters can be expressed as $p(\omega_{1:n}) = p(\omega_1)\prod_{i=2}^n p(\omega_i | \omega_{1:(i-1)} )$. It can be shown that if $\omega_i |G \sim G$ and $G \sim DP(\alpha G_0)$, then each conditional in the product follows a P\'olya Urn Scheme \cite{blackwell1973}

\begin{equation}
\omega_i |  \omega_{1:(i-1)} \sim \frac{\alpha}{\alpha + i - 1} G_0(\omega_i) + \frac{1}{\alpha + i -1 } \sum_{j<i} I(\omega_i = \omega_j) \ \ \text{for $i = 2,\dots, n$}
\end{equation}
where $\omega_1 \sim G_0$. Using this expression, the conditional posterior of $\omega_i$, given $\omega_{1:(i-1)}$, is \cite{muller2015},  

\begin{equation}\label{eq:posterior}
\begin{split}
p(\omega_i | \omega_{1:(i-1)}, D ) & \propto \frac{\alpha}{\alpha + i - 1} p(\omega_i | D_i, G_0)\int_{\omega_i} p(D | \omega_{1:i}) dG_0(\omega_i) + \frac{1}{\alpha + i -1 } \sum_{j<i} p(D_i | \omega_{j}) I(\omega_i = \omega_j) \\
\end{split}						
\end{equation}
where $D_i = ( L_i, A_i, Y_i, Z_i )$ is the data vector for the $i^{th}$ subject. The conditional posterior of each parameter is another P\'olya Urn scheme. The posterior clustering of patients is evident in Equation \ref{eq:posterior}. The parameter $\omega_i$ can equal one of the previously drawn parameters, $\omega_j$, with probability proportional to the subject's likelihood evaluation under $\omega_j$, $(\sum_{j<i}I(\omega_i = \omega_j)) \cdot p(D_i | \omega_j)$. Or, with probability proportional to $\alpha \int_{\omega_i} p(D | \omega_{1:i}) dG_0(\omega_i)$, $\omega_i$ is drawn from the posterior based on a prior draw of $\omega_i$, $p(\omega_i | D_i, G_0)$. \\

If a subject is quite unique so that its likelihood evaluation is low under all of the existing parameters, then it is relatively more likely for this subject to be assigned its own parameters. One can immediately see how this property would be useful for modeling cost data - where some subjects have such uniquely high costs that a better overall fit is achieved by modeling them separately. Finally, note that as $n$ gets large and $i$ approaches $n$, the prior probability $\alpha/(\alpha + i -1)$ of the $i^{th}$ subject being assigned to a new cluster goes to zero. This property helps prevent overfitting.\\

The conditional posterior in Equation \ref{eq:posterior} forms the basis of a Metropolis-in-Gibbs sampler we use to sample $\omega_{1:n}$ from the full posterior, $p(\omega_{1:n}| D) = p(\omega_1 | D )\prod_{i=2}^n p(\omega_i | \omega_{1:(i-1), D} )$. The sampler proceeds in the spirit of Neal's Algorithm 8 \cite{Neal2000} by introducing latent, cluster membership indicators, $c_{1:n}$, for the subjects.\\

We initialize the algorithm by partitioning subjects to one of $K$ initial clusters. In iteration $t$, where we have $K^{(t)}$ occupied clusters indexed by $k$, we draw from the conditional posterior of each model parameter based on the likelihood evaluation from all subjects with $c_i = k$. For parameters with no conjugate priors - such as $\eta_i$ and $\gamma_i$ in Equation \ref{eq:genmod} - we use a Metroplis step \cite{Metropolis1953, Hastings1970}. Conditional on these parameter draws, $\omega_{1:n}^{(t)}$, we update cluster assignment indicators for each subject with
\[ c_i^{(t+1)} | c_{1:(i-1)}^{(t)}  \sim Multinomial\Big(1; \frac{1}{i-1 + \alpha} p(D_i | \omega_1^{(t)}), \dots, \frac{1}{i-1 + \alpha} p(D_i | \omega_{i-1}^{(t)}), \frac{\alpha}{i-1 + \alpha} p(D_{i} | \omega_0 )   \Big)\]

Above, $\omega_0 \sim G_0$ is a draw from the prior. Notice that in each iteration every subject has a $\frac{\alpha}{i-1 + \alpha} p(D_{i} | \omega_0 ) $ probability of being assigned to a new cluster if the proposed parameters, $\omega_0$, yield a higher likelihood evaluation than any of the existing parameters. A more detailed presentation of the algorithm is provided in the appendix.

\subsection{Models Hyperparameters}

The two hyperparameters of the model in Equation \ref{eq:genmod} are the choice of base distribution, $G_0$, and the concentration parameters, $\alpha$. A requirement for the base distribution is that it be over the space of $\omega_i =(\beta_i,\phi_i, \gamma_i, \eta_i, \theta_i)$. Priors for each element of $\omega_i$ can be chosen separately (e.g. a Gaussian prior for $\beta_i$ and a inverse-gamma prior for $\phi_i$, etc). Prior independence may also be assumed - so that $G_0$ is the product over these parameter-specific priors. Conjugate priors for each parameter may be used, if possible, to simplify MCMC computation. \\

The concentration parameter $\alpha$ governs how frequently new clusters appear. It is often described as a prior sample size for a new cluster. Following previous analyses \cite{Roy2018}, we have opted to place a $Gamma(1,1)$ prior on $\alpha$ rather than set it at a particular value. The long tail of this distribution expresses the prior belief that observations likely belong to one of the observed clusters - while also expressing some belief that observations may belong to new, previously unseen clusters. \\

As is typically the case in Bayesian models, prior informativeness impacts clustering in small sample settings. If priors are too tight, then proposed clusters will have parameters similar to existing clusters. If priors are too wide around the data, very unreasonable clusters will be proposed and these will seldom be occupied. Fortunately for the purposes of causal inference, the model estimation is done separately from the post-hoc causal effect computation to be described later. One can fit the model several times with different hyperparameters until one is satisfied with the fit. Once the model is chosen, the causal effects can then be computed afterward. This effectively blinds the analyst to the final causal effect estimate during model tuning.

\subsection{Posterior Mode Clustering in the Presence of Label Switching}
\label{sc:labelswitch}

After a suitable burn-in period, the MCMC procedure outlined in the last section yields $T$ draws from the posterior, $\{\omega_{1:n}^{(t)}, c_{1:n}^{(t)} \}_{1:T}$. Often we may like to cluster patients using the posterior as these clusters represent subgroups of patients that differ in terms of their data distribution. This heterogeneity may be substantively interesting. For example, posterior mode clustering would allow us to summarize these distinct groups in terms of observed characteristics - which may motivate future research.\\

In mixture models, however, posterior mode inference on cluster assignment is complicated by label switching \cite{Rod2014} - the fact that cluster labels $c_{1:n}$ do not have consistent meanings across Gibbs iterations. For example, at iteration $t$, a new cluster, labeled cluster 2, may be proposed and all subjects previously in, say, cluster 1 may be re-assigned to this new cluster. Even though the cluster label has changed from 1 to 2, the cluster still contains the same subjects. Therefore, naively taking the mode of the $T$ cluster indicators, $c_i^{(1)}, \dots, c_i^{(T)}$, for each subject as the posterior mode assignment is ambiguous. \\

To meaningfully cluster subjects based on posterior mode, we perform a deterministic relabeling of cluster indicators \cite{stephens2000}. We compute for each iteration $t$ an $n\times n$ adjacency matrix with a one in the $(i,j)^{th}$ entry indicating patients $i$ and $j$ were clustered together and zero indicating otherwise. The element-wise mean of this matrix across the $t$ iterations gives us a posterior mode matrix showing how often any two patients were clustered together \textit{a posteriori}. To obtain cluster assignments, we select one of the $T$ adjacency matrices that is closest in the $L_2$ sense to the posterior mode matrix. More details regarding the relabeling is provided in the appendix.

\section{Counterfactual Prediction and Estimating Causal Contrasts}
\label{sc:causalderivs}
\subsection{Review of Counterfactuals and Causal Estimation}
We first provide a brief, motivating review of causal estimation before discussing our BNP standardization procedure. Consider observing $D=(Y_i,A_i, L_i)_{i=1:n}$ as defined in Section \ref{sc:genmod}. Using potential outcomes notation \cite{rubin1978, Rubin2005}, let the random variable $Y_i^{A_i=a}$ represent the potential outcome under treatment $A_i=a$. Then the causal effect of $A_i$ on $Y_i$ is given by $Y_i^{A_i=1} - Y_i^{A_i=0}$. While this individual-level causal effect is not identifiable, we can identify the $\textit{marginal}$ causal effect $\Psi = E\Big[Y_i^{A_i=1} - Y_i^{A_i=0}\Big]$ under the following assumptions
\begin{itemize}
	\item Ignorability: $Y_i^{A_i=a} \perp A_i=a | L_i$. Conditional on observed confounders, potential cost is independent of treatment assignment. Unmeasured confounding, for example, would be a violation of this assumption.
	\item Consistency: $Y_i^{A_i=a} = Y_i | A_i =a$. That is, $Y_i$ observed under the actual treatment $A_i=a$ is equal to $Y_i^{A_i=a}$. Non-adherence to treatment assignment is an example of a violation of consistency. 
	\item No interference: $Y_i^{A_i=a} \perp A_j, \ \ \forall i \neq j$. one subject's treatment assignment does not impact another's potential outcome. This assumption may not hold in vaccine studies, for example, where one subject's vaccination status may impact another subject's infection status.
	\item Positivity: $0 < P(A_i = 1| L_i) < 1$. If  $P(A_i = 1| L_i) = 1$, then there is some subpopulation, in terms of $L$, for which we would not have any control subjects for comparison. Mathematically, $Y^0$ would not be defined in this subpopulation.
\end{itemize}
Intuitively, $\Psi$ represents the average difference in the outcome had all patients taken treatment $A=1$ versus $A=0$. Under these assumptions, the method of standardization expresses $E[Y_i^{A=a}]$ in terms of observed data, $D$. Keil et al. \cite{Keil2017} developed a parametric Bayesian approach to the g-formula, which is a generalization of standardization to settings with time-varying treatment/confounding. \\

In this Bayesian approach, inference about the causal effect is conducted using the posterior predictive distribution of the outcome. Let $\tilde{Y}^a$ denote the posterior predictive outcome under intervention $A=a$ with predictive distribution $p(\tilde{Y}^a | D)$. Also, let $\tilde{L}$ denote a posterior predictive draw of confounders. If the causal assumptions hold, the posterior predictive mean under intervention $A=a$ is given by 
\begin{equation} \label{eq:stdpostpredmean}
\begin{split} 
E(\tilde{Y}^a |   D) & = \int_{ \theta} \int_{ \beta} \int_{\tilde{L}} E(\tilde{Y} | \tilde A=a, \tilde{L},  \beta) p(\tilde{L} |  \theta) p( \beta,  \theta |  D) \ d \tilde{L} \ d\beta \  d \theta \\
\end{split}
\end{equation}
Above, $ \beta$ and $ \theta$ are parameter vectors that govern the conditional distribution of the outcome and the distribution of the confounders, respectively.  Similar to frequentist standardization, Equation \ref{eq:stdpostpredmean} simply averages a prediction model for the outcome, $ E(\tilde{Y} | \tilde A, \tilde{L},  \beta)$, over the confounder distribution, $p(\tilde{L} |  \theta)$ , and the posterior distribution of the parameters, $p( \beta,  \theta |  D)$. We can take $E(\tilde{Y}^1 |   D) - E(\tilde{Y}^0 |   D)$ to be a Bayesian point estimate of $\Psi$. Percentiles from the posterior predictive distribution $p(\tilde{Y}^a | D)$ can be used to form credible intervals for $\Psi$. \\

This approach crucially requires both a correctly specified prediction model for the outcome as well as an accurate estimate of the marginal confounder distribution. Thus, two sets of assumptions are required for causal estimation - identifiability assumptions and modeling assumptions. Once they are satisfied, the standardization formula in Equation \ref{eq:stdpostpredmean} essentially acts as formula for predicting the marginal potential outcome $Y^{a}$. However, even if identifiability assumptions are met, the prediction and confounder models are likely to be misspecified. This is especially the case in medical cost data - where multimodality, zero-inflation, and skewness are the norm rather than the exception. Analyzing such data requires flexible models for Equation \ref{eq:stdpostpredmean} that are robust to such pathologies. \\

In the next section, we develop a nonparametric standardization procedure using the DP mixture model proposed in Equation \ref{eq:genmod}. We present the relevant posterior predictive distributions as well as outline Monte Carlo Markov Chain (MCMC) methods to sample form these distributions. We also suggest posterior checks for assessing the validity of positivity - the only testable identification assumption.

\subsection{Sampling from the Posterior Predictive Distribution}
As described previously, Bayesian standardization is done through the posterior predictive distribution. The model outlined in Equation \ref{eq:genmod} yields a flexible posterior predictive distribution, which in turn yields robust causal effect estimates. Under standard causal identification assumptions outlined earlier, the posterior predictive distribution of potential outcome $\tilde Y^{a}$ is given by

\begin{equation} \label{eq:postpreddist}
\begin{split}
p(\tilde Y^a | D ) & =\frac{\alpha}{\alpha + n} \int_{\tilde \omega} \int_{\tilde L}  p(\tilde Y | A=a, \tilde L, \tilde \omega) d P(\tilde L | \tilde \omega) dG_0(\tilde \omega) \\
					& \ \ \ \ \ \ \ \ \ + \frac{1}{\alpha + n }  \int_{\omega_{1:n}} \Big[  \int_{\tilde L} \sum_{j=1}^{n}  p(\tilde Y | A=a, \tilde L, \omega_i) d P(\tilde L | \omega_i)\Big] d P( \omega_{1:n}| D) \\
\end{split}
\end{equation}

Combining the above with the distributional forms from Equation \ref{eq:genmod}, the posterior predictive mean of the potential outcome is,

\begin{equation} \label{eq:postpredmean}
\begin{split}
E(\tilde Y^a | D ) &=  \frac{\alpha}{\alpha + n} \int_{\tilde \omega} \int_{\tilde L} (1 - \pi(\tilde x_a'\tilde \gamma  ) )\cdot \tilde x_a'\tilde \beta \cdot d P(\tilde L | \tilde \theta) dG_0(\tilde \omega) \\
& \ \ \ \ \ \ \ \ \ \ \ \ + \frac{1}{\alpha + n }  \int_{\omega_{1:n}}  \Big[ \int_{\tilde L} \sum_{j=1}^{n} (1 - \pi(\tilde x_a'\gamma_i  ) )\cdot \tilde x_a' \beta_i \cdot d P(\tilde L | \theta_i)\Big] dP( \omega_{1:n}| D) \\
\end{split}
\end{equation}

Above, $\tilde x_a = (1, a, \tilde l )'$. A BNP point estimator of the causal effect, $\Psi$, can be taken to be

\begin{equation}
\hat{\Psi}_{BNP} = E(\tilde Y^1 | D ) - E(\tilde Y^0 | D )
\end{equation}

The integrals in Equations \ref{eq:postpreddist} and \ref{eq:stdpostpredmean} are generally not computable in closed form. However, given draws $( \tilde l^{(t)}, \tilde \omega_{1:n}^{(t)} )_{t=1:T} $ from the posterior distribution $ p( \omega_{1:n}| D) $ and the posterior predictive $p(\tilde l | \omega_i)$, we can obtain $T$ draws from the posterior predictive distribution, $\{\tilde y^a_t \}_{1:T}$. We can use these posterior draws to evaluate the integral in Equation \ref{eq:stdpostpredmean} using Monte Carlo integration as,
\begin{equation} \label{eq:mcmcmean}
\begin{split}
E(\tilde Y^a | D ) & \approx \frac{1}{T} \sum_{t=1}^{T} \frac{\alpha}{\alpha + n} \int_{\tilde \omega} \int_{\tilde L} (1 - \pi(\tilde x_a^{(t)'}\tilde \gamma  ) )\cdot \tilde x_a^{(t)'}\tilde \beta \cdot d P(\tilde L | \tilde \theta) dG_0(\tilde \omega) + \frac{1}{\alpha + n } \sum_{j=1}^{n} (1 - \pi(\tilde x_a^{(t)'}\gamma_i^{(t)}  ) )\cdot \tilde x_a^{(t)'} \beta_i^{(t)} \\
\end{split}
\end{equation}
Above, $\tilde x_a^{(t)} = (1, a, \tilde l^{(t)} )'$ are the regression model vectors under intervention $A=a$ and a draw of confounders $\tilde l^{(t)}$. For the $t^{th}$ draw, the inner integrals in Equation \ref{eq:mcmcmean}  can be evaluated numerically by drawing parameters from the prior $\tilde \omega_0 \sim G_0$, then drawing confounders conditional on these prior draws $\tilde l \sim p(\tilde l | \tilde \omega_0)$. \\

From this equation we can see that our model predicts cost by averaging the predictions from each of the subject-specific models with prior weight $\frac{1}{\alpha + n}$ and a prediction from a potentially new set of parameters with weight $ \frac{\alpha}{\alpha + n}$. In this sense, DP models can - like BART - be viewed as ensemble models. In addition to computing expectations, we can use the posterior predictive draws, $\{\tilde y^a_t \}_{1:T}$, to compute percentiles. These can be used to form credible intervals for $\Psi$. \\

Quantile causal effects and quantile counterfactuals \cite{Xu2018} may also be computed from Equation \ref{eq:postpreddist}. To do this, we estimate the posterior predictive CDF of the potential outcome, $F_a(v) = P(Y^a<v|D)$,  using $\{\tilde y^a_t \}_{1:T}$

\begin{equation} \label{eq:postpredcdf}
\begin{split}
	F_a(v)=P(Y^a\leq v|D) = \int_\infty^v p(\tilde Y^a | D ) & \approx \frac{1}{T} \sum_{t=1}^T I(\tilde y^a_t \leq v ) \\
\end{split}
\end{equation}

We can use the inverse of the estimated CDF to predict quantile potential outcomes (and, therefore, causal effects). For instance, the median causal effect can be estimated as $F^{-1}_1(.5) - F^{-1}_0(.5)$.

\subsection{Assessing the Positivity Assumption}
Positivity is the only identification assumptions that can be assessed empirically. The assumption requires that the probability of treatment is bounded $0<P(A=1| L=l)<1, \ \forall l$. Violations of positivity imply that there are subgroups of the data for which no comparator patients exist - thus forcing the model to extrapolate. Incorrect extrapolation in these regions will bias causal effect estimates. In this section we outline how posterior predictive draws from our proposed model can be used to estimate subject-level propensity scores which in turn can be used to assess positivity. There are many approaches to handling violations once they are identified \cite{Maya2012}, but these are out of scope for this paper and we leave it to future work. Merely providing a framework for assessing this assumption is itself a new contribution to the zero-inflated literature. \\

Under the model in Equation \ref{eq:genmod}, the posterior predictive probability of treatment, conditional on covariates, $l$, is given by 

\begin{equation}
\begin{split}
P(\tilde A | \tilde L=l, D) & = \frac{ \int_{\omega_{1:n}} \big[  \frac{\alpha}{\alpha + n} \int_{\tilde \omega} p(\tilde A| \tilde L, \tilde \omega)p(\tilde L| \tilde \omega) dG_0(\tilde \omega) + \frac{1}{\alpha + n } \sum_{j=1}^{n} p(\tilde A| \tilde L, \omega_j)p(\tilde L| \omega_j )\big] dP(\omega_{1:n}| D)}{ \int_{\omega_{1:n}}   \big[ \frac{\alpha}{\alpha + n}  \int_{\tilde \omega} p(\tilde L| \tilde \omega) dG_0(\tilde \omega)  + \frac{1}{\alpha + n } \sum_{j=1}^{n} p(\tilde L| \omega_j) \big] dP(\omega_{1:n}| D) } \\
\end{split}
\end{equation}

From the generative model, we know the precise forms of all of the involved distributions. Since the integrals are analytically intractable, we evaluate them using Monte Carlo with posterior draws $\omega_{1:n}^{(t)}$ for $t=1, \dots, T$,

\begin{equation}\label{eq:postpredtrt}
\begin{split}
P(\tilde A = 1 | \tilde L=l, D) & \approx \frac{1}{T} \sum_{t=1}^T \frac{ \frac{\alpha}{\alpha + n} \int_{\tilde \omega} p(\tilde A| \tilde L, \tilde \omega)p(\tilde L| \tilde \omega) dG_0(\tilde \omega) + \frac{1}{\alpha + n } \sum_{j=1}^{n} p(\tilde A| \tilde L, \omega_j^{(t)})p(\tilde L| \omega_j^{(t)} ) }{ \frac{\alpha}{\alpha + n}  \int_{\tilde \omega} p(\tilde L| \tilde \omega) dG_0(\tilde \omega)  + \frac{1}{\alpha + n } \sum_{j=1}^{n} p(\tilde L| \omega_j^{(t)}) } \\
\end{split}
\end{equation}

This allows us to estimate $P(\tilde A_i = 1 | \tilde L=l_i, D)$ for each subject, $i$, in our sample. Histograms can be plotted for treated and untreated patients separately. Separated distributions indicate a lack of overlap and, therefore, high posterior likelihood of a positivity violation.

\section{Simulation Study}

In this section we present simulation results evaluating the performance the standardization method using our proposed zero-inflated DP model. Namely, we evaluate bias of $\hat \Psi_{BNP}$, coverage of the corresponding interval estimates, and precision of the estimate as measured by interval width. We compare our method to average treatment effects (ATEs) from existing methods that may be considered by researchers faced with complicated, zero-inflated distributions - namely BART, a doubly robust estimator, and two parametric Gamma models. BART is a nonparametric, tree-based ensemble model for the conditional mean of the outcome. The doubly robust estimator is a two-part model for treatment assignment and the outcome. We use a boosted logistic regressions for the treatment model and a Gaussian model for the outcome. The first parametric model is a Gamma hurdle model - a two-part model that explicitly models the probability of the outcome being zero with a logistic regression, while modeling positive outcomes with a Gamma regression. The second parametric model is a naive, yet somewhat common, approach of adding .01 to zero outcome values and modeling this transformed outcome using a Gamma regression. We refer to this as the Gamma $+.01$ model.\\

We simulate from two data generating processes (DGPs). In a clustered DGP, we simulate data from three distinct clusters - each with its own set of parameters that governs confounder distributions, treatment assignment, zero-inflation, and Gamma-distributed positive outcomes. The Gamma distribution is used to simulate realistic cost data that are non-negative and skewed within each cluster. Thus, the \textit{local} conditional outcome distribution assumed in Equation \ref{eq:genmod} is deliberately misspecified. The average proportion of zero outcomes in each dataset was about $45\%$ - a relatively extreme setting but not uncommon in actual cost data. \\

In a parametric DGP, we simulate data from a single cluster with a common covariate distribution, treatment assignment model, zero-inflation, and Gamma-distributed positive outcomes. The data is still skewed, but not multimodal. For each DGP, we simulate 1000 datasets with 3000 subjects each. We simulate with one continuous covariate and four binary covariates, all of which affect zero-inflation, treatment probability, and the outcome. In this setting, the average proportion of zeros in each dataset was about $55\%$ \\

\input{SimSetting_MultiVar.tex}

In the clustered setting, the zero-inflated DP model produces effect estimates with the smallest bias - $-8.1\%$ of the true value with close to nominal coverage of $94.2\%$ - due to it's ability to capture both skewness, structural zeros, and multimodality. The small bias may be attributed to the local misspecification of the model. The Gamma hurdle model has the second best performance in terms of bias - a consequence of its inability to capture multimodality. Nevertheless, it performs better than BART and the double robust estimators likely due to its explicit modeling of structural zeros and skewness. The former two models capture neither zero-inflation nor multimodality and consequently perform poorly. \\

In the parametric setting, the zero-inflated DP model again exhibits low bias and close to nominal coverage as well. BART and the doubly robust models have lower bias, but exhibit slight overcoverage (about 96\%) in the interval estimates. The Gamma hurdle model is correctly specified in the parametric DGP and so performs the best - exhibiting the lowest bias of 1.4\% as well as $95.1\%$ coverage and yielding the shortest interval. Relative to the correctly specified model, the Zero-inflated DP has only a slightly wider interval length on average (22034.1 versus 21778.7) - suggesting minimal efficiency loss. BART and the doubly robust estimators both have wider intervals than the DP. \\

The particularly bad performance of the naive Gamma +.01 model - under both DGPs - should be noted. While it is a simple, seemingly harmless trick, adding a small constant severely degrades the accuracy and precision of treatment effect estimates. \\

The relatively good performance of the zero-inflated DP model is due to its data-adaptive nature. It introduces more parameters if the data distribution is complicated, yet shrinks towards the local model if the data distribution is simple. This allows for quality treatment effect estimates under both simple and pathological data distributions - with minimal efficiency loss if the parametric model is correct. In the next section, we highlight the proposed model's utility in not only producing quality ATE estimates, but also in capturing the full joint distribution and patient clustering.

\section{Application: Inpatient Medical Costs for Endometrial Cancer}

In this section, we use the proposed DP mixture of zero-inflated regressions to analyze inpatient costs among patients with endometrial cancer. Patients who were diagnosed with endometrial cancer between 2000 and 2014 were identified in Medicare's Surveillance, Epidemiology, and End Results (SEER) database. Patients who were assigned to either radiation or chemotherapy at diagnosis were followed for a maximum of two years. The total inpatient costs, measured in 2018 US dollars, accrued over the followup period was recorded and is our primary outcome of interest. Inpatient costs are costs that accrue during overnight hospitalizations - they do not include costs such as prescription treatment costs, outpatient costs, or hospice care costs.\\
\input{bsl_table.tex}
Table \ref{tbl:bsl} presents baseline characteristics of the two treatment groups. There is a significant proportion of zero costs - 15.2\% in the chemotherapy arm versus 7.9\% in the radiation arm. Chemotherapy subjects have lower inpatient costs over the follow up period. However, there may be several confounding factors. The Chemotherapy arm has a higher average annual household income, lower proportion of white patients, and lower rank on the Charlson Commorbidity Index (CCI). The standardized mean difference for all of these variables is $>.1$. The two groups also differ based on baseline tumor grade and stage. The chemotherapy arm contains a higher proportion of patients at stages I-N0 and I-A - which are the lowest (less severe) stages. \\

In the following subsections, we will demonstrate how our method can be used to model the data from several angles and produce various useful outputs. We can simultaneously identify subgroups of patients with distinct cost-covariate profiles, construct predictive distributions that accurately fit the complex data, and compute various causal treatment contrasts. All subsequent results are from posterior sampling of the model in Equation \ref{eq:genmod}. We control for race, CCI, household income, cancer grade and stage in both the positive outcome model and the zero-probability model. We model treatment assignment as a function of these confounders as well. Since our model is generative, we assume Gaussian distributions for CCI and household income and Bernoulli distributions for binary covariates. More details about hyperparameter choices and posterior sampling results are provided in the appendix.

\subsection{Multi-modality and Clustering Results}
This particular dataset contains a very heterogeneous set of patients in terms of their observed costs and covariates. Some have extremely high costs, more comorbidities, and come from varying socio-economic backgrounds. As outlined before, the Dirichlet process prior induces a clustering of patients who are similar in terms of these observed characteristics - which can help identify these heterogeneous subgroups. Since the DP mixture assumes there are infinitely many clusters,  we can cluster patients nonparametrically rather than having to specify the number of clusters beforehand.\\
\begin{figure}[H] 
	\centering 
	\includegraphics[scale=.45]{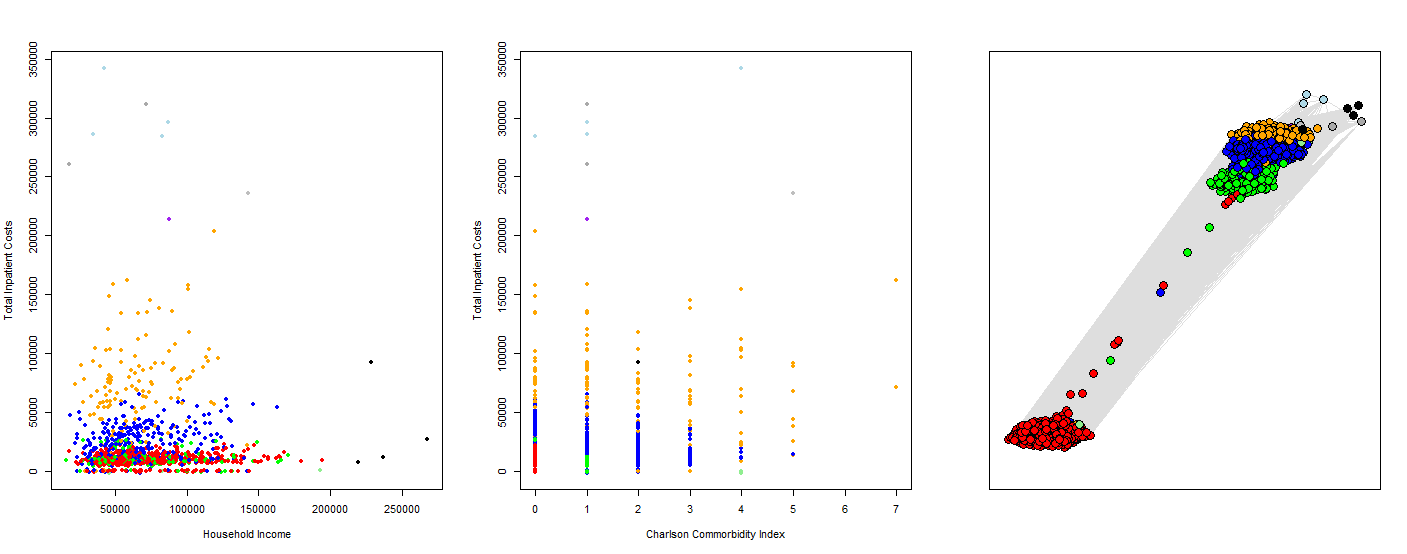}
	\caption{\small Clustering results from the zero-inflated mixture. Our model identifies six clusters - colors indicate posterior mode cluster membership. Each cluster has its own distinct cost-covariate profile.}
	\label{fig:dataclust}
\end{figure}
Two important confounders of costs and treatment assignment are household income and CCI. Patients from more affluent backgrounds may opt for different treatments and likely be willing to pay different amounts. Similarly, patients with more comorbidities may be assigned to different treatments and be more costly in general. The first two panels of Figure  \ref{fig:dataclust} visualize cost along these important dimensions. As can be seen, our model detects several clusters, each displayed with its own color representing posterior mode cluster assignment. While we initialize the model with five clusters, the model identified ten clusters in the posterior - introducing five additional clusters to accommodate the complexity of the data. Focusing on the first panel, we see the orange cluster has very high costs, the blue cluster has moderately high costs, while the green and red clusters have lower costs. There are two interesting things to note. First, the light blue and gray clusters represent patients who have such distinctly high costs that the DP model places them in their own cluster. Second, the black points represent patients who, while having similar costs to most patients, have distinctly high household income. Thus the DP model places them in their own cluster. From this black cluster we can see clearly that the clustering is happening in multiple dimensions rather than only on the cost space. \\

Similarly, we cannot see much difference between the green and red clusters on the cost-household income space. However, the second panel shows that these patients occupy distinct places on the cost-CCI space - with the red cluster ranking lower than green on CCI. It may be clear at this point that quality of cluster visualization in two-dimensions is limited by the need to choose the variables on each dimension. The third panel solves this issue by visualizing the entire posterior mode adjacency matrix discussed in Section \ref{sc:labelswitch} as a network diagram. In this dimension-free visualization, each node represents a patient and vertices connecting two patients have a length inversely proportional to how often they were clustered \textit{a posteriori}. Patients close together were clustered frequently together. We can also get a sense of the uncertainty around these cluster assignments using this diagram. For example, the nodes between the red and green cluster have very uncertain assignment. About half the time, they were clustered with the red patients and the other half they were clustered with the green patients. To gain a deeper understanding of these clusters, we can summarize observed characteristics of patients by posterior mode assignment, as is done in Table \ref{tbl:clust} (we omitted some of the smaller clusters for compactness). The columns are sorted from lowest to highest cost. 

\input{clust_table.tex}

In the table, we can get a better sense of the composition of the orange cluster of relatively high-cost patients. Average cost in this cluster was \$71,139. We can see that the distribution of CCI in this group is skewed much higher. These patients have significant comorbidities at baseline. Such tables can be useful for both providing descriptive insight and motivating future research.

\subsection{Cost Prediction in Presence of Zero-Inflation}
\begin{figure}[H] 
	\centering 
	\includegraphics[scale=.5]{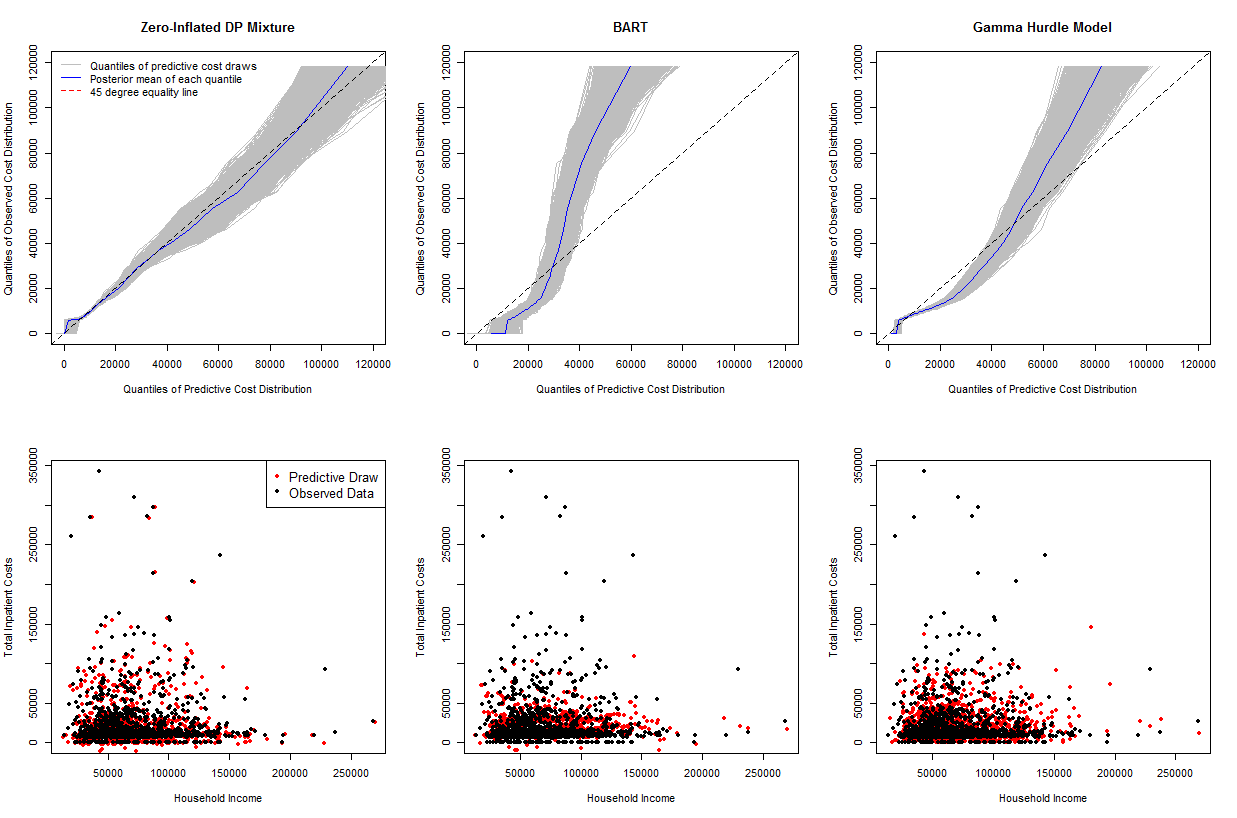}
	\caption{\small Top row: Percentiles (.02 - .98 in increments of .02) of the observed cost distribution against predictive cost distributions from various models. Blue line represents the posterior mean of each percentile. The dotted line indicates equality. The DP mixture opens new cluster to capture skewness - resulting in a predictive distribution closely matching the observed data. The BART model and hurdle model cannot capture skewness. This is also demonstrated in the bottom row: the DP model occasionally predicts very high costs, while predictions from BART and hurdle models hardly ever predict such high costs.}
	\label{fig:qqplots}
\end{figure}

Induced clustering is the core strength of DP mixtures: a single parametric model estimated using heterogeneous dataset will have worse fit than several local parametric models fit on more homogeneous partitions. The clustering not only accounts for multimodality, but also allows the model to capture skewness. Figure \ref{fig:qqplots} demonstrates the proposed model's effectiveness at capturing the cost distribution. The first row shows QQ plots of the observed cost quantiles against quantiles of the predictive cost distribution. Each gray line is a draw of the same size as the data from the predictive cost distribution. The blue line indicates the mean of each percentile across these predictive draws while the dashed line indicates equality (a perfect fit). The predictive cost distributions from the DP mixture oscillate around the 45 degree line. This is not the case for the BART and hurdle models. The predictive distributions are all far above the 45 degree line - indicating that the predictive draws are consistently lower than observed costs. Predictive draws from these models are not faithful representations of the observed costs - though the hurdle model does a considerably better job than BART. The second row of the figure provides some insight into the performance difference. 
The DP model occasionally predicts very high costs, while having the bulk of the predictions at $<\$50,000$. Both BART and the Hurdle model capture the lower end of the cost distribution well - also predicting the bulk of the costs at $<\$50,000$. However, they rarely predict costs at the high end - thus, failing to capture skewness.

\subsection{Estimating Causal Contrasts and Assessing Overlap}

Finally, we use our method to estimate differences in costs that would have accumulated under hypothetical interventions where everyone was assigned to radiation versus chemotherapy. We apply the standardization procedure outlined in Section \ref{sc:causalderivs} to compute an average causal effect and a median causal effect. Moreover, we compute a risk ratio contrasting the probability of having zero costs under radiation versus chemotherapy. Posterior means and credible intervals are displayed in Table 4. Under standard causal identification assumptions, we estimate the causal difference in costs to be $1672.62$ - showing radiation therapy to be more expensive. We estimate a median causal difference to be $872.68$. Finally, we estimate that the probability of having zero costs under radiation therapy is 50\% lower than under chemotherapy. This results are largely consistent with unadjusted results (see Table \ref{tbl:bsl}). \\

Average treatment effects from BART and the Gamma hurdle model are also presented for reference. These were computed by averaging posterior predictive difference in costs (between radiation and chemotherapy) over the empirical confounder distribution. They are roughly in-line with the DP mixture estimates but suffer from relative ineffectiveness at predicting high costs, as explained in the previous section. The risk ratio estimate from the Gamma hurdle model is similar to the DP model estimate. \\

In addition, we compare our results to the Gamma +.01 model. This average causal effect estimate differs greatly from the other three models - consistent with simulation results.
\input{DataRes_CausalEsts.tex}
Causal interpretations of these quantities are conditional on the causal assumptions being met. While, we can never be certain of ignorability, we can empirically assess positivity - a crucial check when making causal inferences. \\
\begin{figure}[H] 
	\centering 
	\includegraphics[scale=.45]{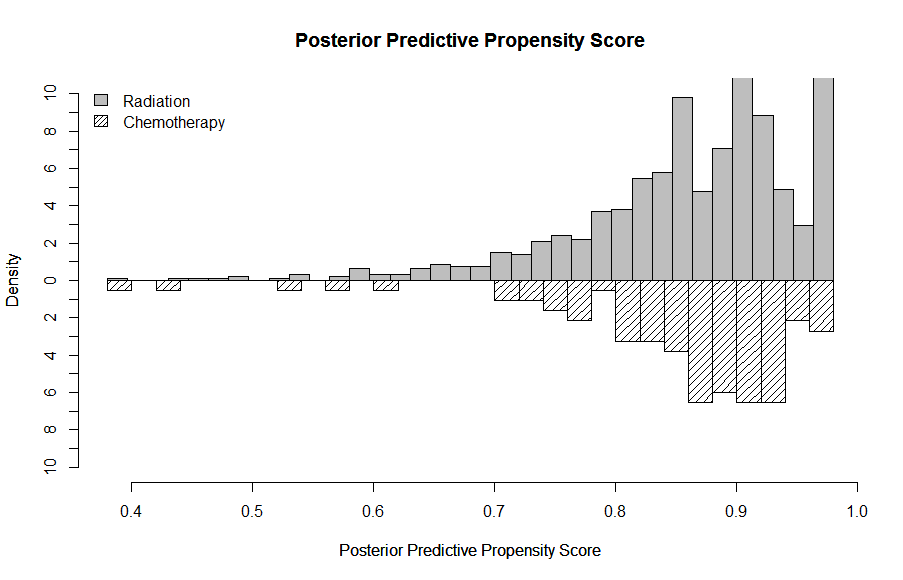}
	\caption{\small Posterior propensity scores for both groups indicate adequate overlap - suggesting positivity violations are minimal.}
	\label{fig:pscore}
\end{figure}
Figure \ref{fig:pscore} plots the distribution of the posterior mean propensity scores among both arms obtained through Equation \ref{eq:postpredtrt}. There is adequate overlap between the two treatment groups and we should not be too concerned about positivity violations. Since there are many more radiation patients than chemotherapy patients, both distributions are close to 1 (the indicator of radiation treatment).

\section{Discussion and Future work}

The proposed DP mixture is ideal for capturing joint distributions with continuous, zero-inflated outcomes. It is multipurpose: simultaneously modeling structural zeros, inducing clustering to handle multi-modality, and accommodating skewness in the outcome and covariates. As was shown in the data analysis portion, these traits give the DP model a competitive edge over the nonparametric BART model and parametric Gamma hurdle model when it comes to capturing cost distributions and estimating causal effects. As shown in simulations, the incorporation of this model into a Bayesian standardization procedure yields low-bias point estimates and interval estimates with close to nominal coverage even in highly pathological data generating settings. In parametric settings, the model maintains this performance with less efficiency loss relative to comparator methods. At the same time, posterior draws can be used to perform posterior checks evaluating the validity of positivity - an important check when estimating causal effects. \\

One might expect BART to perform better than it did in simulations and data analysis. After all, BART is also ``effectively nonparametric"\cite{chipman2010}. Just as DP mixtures partition data into homogeneous clusters and model them with cluster-specific parameters, BART partitions data into homogeneous terminal nodes via the node-splitting mechanism of its trees. Just as BART forms predictions by ensembling over individual trees, DP mixtures form predictions by ensembling over cluster-specific models. BART's relatively poor performance in these settings likely stem for the fact that it is a conditional model for the mean of a \textit{Gaussian} outcome. While it flexibly models the mean, it still assumes that the error term is Gaussian - yielding biased estimates in both simulation settings where the data are drawn from a skewed distribution like the Gamma. Moreover, it does not account for multi-modality as it assumes the data are generated from a single mean function and single variance. The DP mixture makes no such assumptions. Moreover, since BART proposes trees randomly, it is unlikely that subjects at the high end of the outcome distribution will have their own terminal tree node. The DP mixture, by contrast, is likely to propose new clusters that accommodate such subjects. This allows for not only robust treatment effect estimation, but also better estimation of the entire outcome distribution, as was shown in Figure \ref{fig:qqplots}. We note that the currently unpublished work of George et al \cite{george2018} extends BART by modeling the error term nonparametrically using a DP mixture. This may better equip BART to handle skewness, though multi-modality will likely still be a significant modeling challenge.   \\

Finally, we consider several extensions alluded to throughout the paper. First, while our model provides a framework for assessing positivity, designing a solution within the framework of our model is an important goal and we leave it to future work. In terms of causal effect estimation, we note that standardization can only account for time-constant treatment and confounding. In our cost data analysis, for example, patients may switch between radiation and chemotherapy throughout the follow-up period in ways that depend on (hopefully) observed confounders. Therefore the treatment effects in our data analysis likely estimate intent-to-treat effects rather than causal effects. Estimating causal effects in such settings requires extending our model to handle time-varying confounding. Finally, in the proposed model we specify a local Gaussian for the conditional positive outcome distribution. This yields computational benefits by allowing us to exploit conjugacy, but comes at the cost of ignoring the non-negative nature of cost outcomes in the data analysis. We view this as a good trade-off and note that a more accurate local distribution for positive outcomes would likely only improve the results presented. Since this is a more general purpose method with costs being a motivating applied example, we stress that researchers analyzing different outcomes may consider using other local distribution, but it seems the Gaussian is a good choice as a first-pass.
 
\newpage
\section*{Acknowledgments}
\begin{itemize}
	\item We used the linked SEER-Medicare database and acknowledge the efforts of the Applied Research Program; National Cancer Institute; Office of Research, Development and Information; Centers for Medicare and Medicaid Services; Information Management Services; and SEER program tumor registries in the creation of the SEER-Medicare database.
	\item This work was supported in part by Grant R01GM112327 from National Institute Of General Medical Sciences and Grant 124268-IRG-78-002-35-IRG from the American Cancer Society.
	\item We thank Dr. Emily M. Ko for access and clinical guidance with the data analysis component of this paper.
\end{itemize}

\newpage
\bibliographystyle{unsrt}
\bibliography{ZeroInf} 

\newpage
\appendix

\section{Derivations of Posterior Distributions}

\subsection{Posterior of Parameters}

\begin{equation}
\begin{split}
p(\omega_i | \omega_{1:(i-1)}, D ) & \propto p(D | \omega_{1:i}) p(\omega_i| \omega_{1:(i-1)}) \\
& \propto \frac{\alpha}{\alpha + i - 1} p(D | \omega_{1:i})G_0(\omega_i) + \frac{1}{\alpha + i -1 } \sum_{j<i} p(D_j | \omega_{j}) I(\omega_i = \omega_j) \\
& \propto \frac{\alpha}{\alpha + i - 1} p(\omega_i | D_i, G_0)\int_{\omega_i} p(D | \omega_{1:i}) dG_0(\omega_i) + \frac{1}{\alpha + i -1 } \sum_{j<i} p(D_j | \omega_{j}) I(\omega_i = \omega_j) \\
\end{split}						
\end{equation}

The second line follows by substituting in the P\'olya Urn scheme for $p(\omega_i| \omega_{1:(i-1)})$.  The last line follows from the fact that $p(\omega_i | D_i, G_0) = \frac{p(D | \omega_{1:i})G_0(\omega_i)}{\int_{\omega_i} p(D | \omega_{1:i}) dG_0(\omega_i)}  $.

\subsection{Posterior Predictive Distribution of Potential Outcome}

Causal inference using standardization is based off the posterior predictive mean of the outcome under some intervention $A=a$. Denote this as $\tilde y^a$. Letting tildes denote posterior predictive draws throughout, 

\begin{equation}
\begin{split}
p(\tilde y^a | D ) & =  \int_{\omega_{1:n}}\int_{\tilde l} \int_{\tilde \omega}  p(\tilde y^a | \tilde l, \tilde \omega ,\omega_{1:n} D) p(\tilde l | \tilde \omega ,\omega_{1:n} D) p(\tilde \omega | \omega_{1:n}) p( \omega_{1:n}| D) \ d \tilde \omega \ d \tilde L \ d \omega_{1:n}  \\
\end{split}
\end{equation}

Conventionally, we assume that that $\tilde y^a \perp \omega_{1:n}, D | \tilde \omega$ and $\tilde l \perp \omega_{1:n}, D | \tilde \omega$. That is, conditional on new parameter draws, the new outcome draw is independent of previous observations and their parameters.

\begin{equation}
\begin{split}
p(\tilde y^a | D ) & =  \int_{\omega_{1:n}} \int_{\tilde l} \int_{\tilde \omega}  p(\tilde y^a | \tilde l, \tilde \omega) p(\tilde l | \tilde \omega) p(\tilde \omega | \omega_{1:n}) p( \omega_{1:n}| D) \ d \tilde \omega \ d\tilde L \ d\omega_{1:n}  \\
\end{split}
\end{equation}

Assuming ignorability and consistency hold, 

\begin{equation}
\begin{split}
p(\tilde y^a | D ) & =  \int_{\omega_{1:n}} \int_{\tilde l} \int_{\tilde \omega}  p(\tilde y | A=a, \tilde l, \tilde \omega) p(\tilde l | \tilde \omega) p(\tilde \omega | \omega_{1:n}) p( \omega_{1:n}| D) \ d \tilde \omega \ d\tilde L \ d\omega_{1:n} \\
\end{split}
\end{equation}

Recall that $ \omega_i | \omega_{1},\dots, \omega_{i-1}$  from the P\'olya Urn \cite{blackwell1973}
\[ \omega_i | \omega_{1:(i-1)} \sim \frac{\alpha}{\alpha + i - 1} G_0(\omega_i) + \frac{1}{\alpha + i -1 } \sum_{j=1}^{i-1} I(\omega_i = \omega_j) \]

Substituting $i=n+1$ yields 
\[ \tilde \omega | \omega_{1:n} \sim \frac{\alpha}{\alpha + n} G_0(\tilde \omega) + \frac{1}{\alpha + n } \sum_{j=1}^{n} I(\tilde \omega = \omega_j) \]
Substituting this yields,

\begin{equation}
\begin{split}
p(\tilde y^a | D ) & =  \int_{\omega_{1:n}} \int_{\tilde l} \int_{\tilde \omega}  p(\tilde y | A=a, \tilde l, \tilde \omega) p(\tilde l | \tilde \omega) \Big[  \frac{\alpha}{\alpha + n} G_0(\tilde \omega) + \frac{1}{\alpha + n } \sum_{j=1}^{n} I(\tilde \omega = \omega_j) \Big] p( \omega_{1:n}| D) \ d \tilde \omega \ d\tilde L \ d\omega_{1:n} \\
& =  \int_{\omega_{1:n}} \int_{\tilde l}\Big[ \int_{\tilde \omega}  p(\tilde y | A=a, \tilde l, \tilde \omega) p(\tilde l | \tilde \omega) \frac{\alpha}{\alpha + n} G_0(\tilde \omega)d \tilde \omega  \\
& \ \ \ \ + \frac{1}{\alpha + n } \sum_{j=1}^{n}  \int_{\tilde \omega}  p(\tilde y | A=a, \tilde l, \tilde \omega) p(\tilde l | \tilde \omega)I(\tilde \omega = \omega_j) d \tilde \omega \Big] p( \omega_{1:n}| D) \ d\tilde L \ d\omega_{1:n} \\
& = \int_{\omega_{1:n}} \int_{\tilde l}\Big[  \frac{\alpha}{\alpha + n} \int_{\tilde \omega}  p(\tilde y | A=a, \tilde l, \tilde \omega) p(\tilde l | \tilde \omega) G_0(\tilde \omega)d \tilde \omega + \frac{1}{\alpha + n } \sum_{j=1}^{n}  p(\tilde y | A=a, \tilde l, \omega_i) p(\tilde l | \omega_i)\Big] p( \omega_{1:n}| D) \ d\tilde L \ d\omega_{1:n} \\
\end{split}
\end{equation}

Now the posterior predictive mean is
\begin{equation}
\begin{split}
E(\tilde y^a | D ) & = \int_{\omega_{1:n}} \int_{\tilde l}\Big[  \frac{\alpha}{\alpha + n} \int_{\tilde \omega}  E(\tilde y | A=a, \tilde l, \tilde \omega) p(\tilde l | \tilde \omega) G_0(\tilde \omega)d \tilde \omega \\
& \ \ \ \ \ + \frac{1}{\alpha + n } \sum_{j=1}^{n}  E(\tilde y | A=a, \tilde l, \omega_i) p(\tilde l | \omega_i)\Big] p( \omega_{1:n}| D) \ d\tilde L \ d\omega_{1:n} \\
& = \int_{\omega_{1:n}} \int_{\tilde l}\Big[  \frac{\alpha}{\alpha + n} \int_{\tilde \omega} (1 - \pi(\tilde x'\tilde \gamma  ) )\cdot \tilde x'\tilde \beta \cdot p(\tilde l | \tilde \omega) G_0(\tilde \omega)d \tilde \omega \\
& \ \ \ \ \ + \frac{1}{\alpha + n } \sum_{j=1}^{n} (1 - \pi(\tilde x'\gamma_i  ) )\cdot \tilde x' \beta_i \cdot p(\tilde l | \omega_i)\Big] p( \omega_{1:n}| D) \ d\tilde L \ d\omega_{1:n} \\
\end{split}
\end{equation}

Let $( \tilde l^{(t)}, \tilde \omega_{1:n}^{(t)} )_{t=1:T} $ we have $T$ draws from the posterior distribution, $ p( \omega_{1:n}| D) $ and the posterior predictive $p(\tilde l | \omega_i)$, we can evaluate this integral using Monte Carlo as,
\begin{equation}
\begin{split}
E(\tilde y^a | D ) & \approx  \frac{1}{T} \sum_{t=1}^{T}  \frac{\alpha}{\alpha + n} \int_{\tilde \omega} \int_{\tilde l} (1 - \pi(\tilde x_a^{'} \tilde \gamma  ) ) \cdot \tilde x' \tilde\beta \cdot p(\tilde l | \tilde \omega) G_0(\tilde \omega) \ d\tilde l \ d \tilde \omega  + \frac{1}{\alpha + n } \sum_{j=1}^{n} (1 - \pi(\tilde x_a^{(t)'}  \gamma_i^{(t)}  ) ) \cdot \tilde x' \beta_i^{(t)} \\
\end{split}
\end{equation}
Above, $\tilde x_a = (1, a, \tilde l )'$ and $\tilde x_a^{(t)} = (1, a, \tilde l^{(t)} )'$ are the regression model vectors under intervention $A=a$. For the $t^{th}$ draw, the inner integral can be evaluated numerically by drawing parameters from the prior $\tilde \omega_0 \sim G_0$, then drawing confounders conditional on these prior draws $\tilde l \sim p(\tilde l | \tilde \omega_0)$. \\

\subsection{Posterior Predictive Probability of Treatment}

We can use the model to estimate the propensity score for each subject $i$ using the posterior predictive probability of treatment conditional on subject $i$'s covariates.

\begin{equation}
\begin{split}
P(\tilde A | \tilde L=l, D) & = \frac{p(\tilde A, \tilde L | D)}{p(\tilde L| D)} \\
& = \frac{ \int_{\tilde \omega} \int_{\omega_{1:n}}  p(\tilde A, \tilde L, \tilde \omega, \tilde \omega_{1:n}| D) \ d \tilde \omega \ d \omega_{1:n}}{ \int_{\tilde \omega} \int_{\omega_{1:n}}  p(\tilde L, \tilde \omega, \tilde \omega_{1:n}| D) \ d \tilde \omega \ d \omega_{1:n}} \\
& = \frac{ \int_{\tilde \omega} \int_{\omega_{1:n}}  p(\tilde A| \tilde L, \tilde \omega)p(\tilde L| \tilde \omega) p(\tilde \omega| \omega_{1:n}) p(\omega_{1:n}| D)\ d \tilde \omega \ d \omega_{1:n} }{ \int_{\tilde \omega} \int_{\omega_{1:n}}  p(\tilde L| \tilde \omega) p(\tilde \omega| \tilde \omega_{1:n}) p(\tilde \omega_{1:n}| D)\ d \tilde \omega \ d \omega_{1:n}} \\
\end{split}
\end{equation}

Again, substituting the P\'olya Urn distribution,

\begin{equation}
\begin{split}
P(\tilde A | \tilde L=l, D) & = \frac{ \int_{\tilde \omega} \int_{\omega_{1:n}}  p(\tilde A| \tilde L, \tilde \omega)p(\tilde L| \tilde \omega) \big[ \frac{\alpha}{\alpha + n} G_0(\tilde \omega) + \frac{1}{\alpha + n } \sum_{j=1}^{n} p(\tilde A| \tilde L, \tilde \omega)p(\tilde L| \tilde \omega)I(\tilde \omega = \omega_j) \big] p(\omega_{1:n}| D) \ d \tilde \omega \ d \omega_{1:n}}{ \int_{\tilde \omega} \int_{\omega_{1:n}}  p(\tilde L| \tilde \omega) \big[ \frac{\alpha}{\alpha + n} G_0(\tilde \omega) + \frac{1}{\alpha + n } \sum_{j=1}^{n} I(\tilde \omega = \omega_j) \big] p(\tilde \omega_{1:n}| D) \ d \tilde \omega \ d \omega_{1:n}} \\
& = \frac{ \int_{\omega_{1:n}} \big[  \frac{\alpha}{\alpha + n} \int_{\tilde \omega} p(\tilde A| \tilde L, \tilde \omega)p(\tilde L| \tilde \omega) G_0(\tilde \omega) \ d \tilde \omega + \frac{1}{\alpha + n } \sum_{j=1}^{n} p(\tilde A| \tilde L, \omega_j)p(\tilde L| \omega_j )\big] p(\omega_{1:n}| D) \ d \omega_{1:n}}{ \int_{\omega_{1:n}}   \big[ \frac{\alpha}{\alpha + n}  \int_{\tilde \omega} p(\tilde L| \tilde \omega)G_0(\tilde \omega) \ d \tilde \omega + \frac{1}{\alpha + n } \sum_{j=1}^{n} p(\tilde L| \omega_j) \big] p(\tilde \omega_{1:n}| D) \ d \omega_{1:n}} \\
\end{split}
\end{equation}

Again, given $T$ posterior draws $\omega_{1:n}^{(t)}$ indexed by $t$, we can perform a Monte Carlo evaluation of the integral

\begin{equation}
\begin{split}
P(\tilde A =1 | \tilde L=l, D) & \approx \frac{1}{T} \sum_{t=1}^T \frac{\frac{\alpha}{\alpha + n} \int_{\tilde \omega} p(\tilde A=1| \tilde L=l, \tilde \omega)p(\tilde L=l| \tilde \omega) G_0(\tilde \omega) \ d \tilde \omega + \frac{1}{\alpha + n } \sum_{j=1}^{n} p(\tilde A=1| \tilde L=l, \omega_j^{(t)})p(\tilde L=l| \omega_j^{(t)} )}{ \frac{\alpha}{\alpha + n}  \int_{\tilde \omega} p(\tilde L=l| \tilde \omega)G_0(\tilde \omega) \ d \tilde \omega + \frac{1}{\alpha + n } \sum_{j=1}^{n} p(\tilde L=l| \omega_j^{(t)})} \\
\end{split}
\end{equation}

\section{Metropolis-in-Gibbs Sampler and Relabeling Strategy}

To sample from the model in Equation \ref{eq:genmod}, we use the Metropolis-in-Gibbs sampler outline in Section \ref{sc:postsample}. We describe the MCMC algorithm in more detail. First, introduce latent cluster indicators for the $n$ subjects at iteration $t$ of the algorithm, $c_{1:n}^{(t)}$. In this iteration, each $c_i$ may take on one of $K^{(t)}$ unique values. Let $\mathcal{K}^{(t)}$ be the set of unique cluster labels at iteration $t$. Associated with each of these clusters is a set of cluster specific parameters $\omega_k^{(t)} = (\gamma_k^{(t)}, \beta_k^{(t)}, \phi_k^{(t)}, \eta_k^{(t)}, \theta_k^{(t)})$. \\

The MCMC procedure alternates between updating the cluster-specific parameters, $\omega_k$, conditional on $c_{1:n}$. Then updates $c_{1:n}$ conditional on $\omega_k$. The procedure is given in Algorithm 1.

\begin{algorithm}[H] 
	\caption{Metropolis-in-Gibbs Posterior Sampler for Zero-inflated DP Mixture}
	\begin{algorithmic}[1]
		\State Initialize $c_{1:n}^{(0)}$ to $K^{(0)}$ initial clusters with unique labels $\mathcal{K}^{(0)}$.
		\State Initialize parameters $\omega_k^{(0)}$ for each of these clusters.
		\For{t=1:T} 
		\State Update Cluster-specific Parameters
			\For{$k$ in $\mathcal{K}^{(t-1)}$ }
			\State $\beta^{(t)}_k \sim p(\beta | \phi_k^{(t-1)}, D) \propto \prod_{i| y_i >0} N(y_i| x_i'\beta,  \phi_k^{(t-1)},  c_{1:n}=k ) \cdot G_0(\omega)$
			\State $\phi^{(t)}_k \sim p(\phi | \beta_k^{(t)}, D) \propto \prod_{i| y_i >0} N(y_i| x_i'\beta_k^{(t)}, \phi,  c_{1:n}=k ) \cdot G_0(\omega)$
			\item[]
			\State $\theta_k^{(t)} \sim p(\theta_k | D) \propto \prod_{i} p(l_i | \theta_k, c_{1:n}=k )\cdot G_0(\omega)$
			\item[]
			\State $  \eta_k^{(t)} \sim p(\eta_k|D) \propto \prod_i Ber(A_i| expit(m'\eta_k), c_{1:n}=k ) \cdot G_0(\omega) $
			\State $  \gamma_k^{(t)} \sim p(\gamma_k|D) \propto \prod_i Ber(z_i| expit(x'\gamma_k), c_{1:n}=k ) \cdot G_0(\omega)$
			\EndFor
		\State Conditional on $\omega_k^{(t)}$ for all $k \in \mathcal{K}^{(t-1)} $, update $c_{1:n}$
		\State Propose a new cluster with parameters drawn from the base distribution, $\omega_{new} \sim G_0$
			\For{i=1:n}
			\item[] Update to existing cluster...
			\State $P(c_i = k | c_{-i}, \{\omega_k^{(t)} : \forall k \in \mathcal{K}^{(t-1)}\} , D) \ \propto p(D_i| \omega_{k}^{(t)} ) \cdot \frac{\sum_{j<i} I(c_i=c_j)}{\alpha + n}\ $
			\item[]...or to the newly proposed cluster.
			\State $P(c_i \neq k,  \forall k\in \mathcal{K}^{(t-1)} \ | \ c_{-i}, \{\omega_k^{(t)} : \forall k \in \mathcal{K}^{(t-1)}\}, D) \ \propto p(D_i| \omega_{new} ) \cdot \frac{\alpha}{\alpha + n}\ $ 
			\EndFor

		\EndFor
	\end{algorithmic}
\end{algorithm}
In practice we assume prior independence, so that $G_0=p(\beta|\mu_\beta)p(\phi|\mu_\phi)p(\theta | \mu_\theta) p(\eta | \mu_\eta )p(\gamma | \mu_\gamma )$. Here, the $\mu$ terms represent hyperparameters. In this setting, the prior distribution factors so that in line 6, for example, $G_0(\omega) \propto p(\beta|\mu_\beta) $. Furthermore, we could choose $ p(\beta|\mu_\beta)$ to be a multivariate Gaussian with hyper mean vector and covariance matrix $\mu_\beta = ( \lambda, \Sigma )$. This allows us to performing the sampling in line 6 using conjugacy. The idea is the same for the covariate model update in line 8 - where we could specify independent beta priors for binary variables, normal priors for continuous covariates, and Dirichlet priors for categorical variables. More complicated distributions can be chosen to better model correlations between the parameters at the expense of computational complexity.  \\

The hyperparameters must be chosen with care. If the priors are too widely centered around the data, the clusters proposed from the priors will seldom get occupied and new clusters will not form. For this reason, we recommend calibrating priors empirically. For example, if we place a conditionall conjugate prior on $\beta_k$ as in the example of the last paragraph, it would be sensible to choose $\lambda$ to be the OLS estimate using non-zero outcomes. \\

Since there are no conjugate priors for the logistic models, the updates in lines 9 and 10 are done using a Metropolis step. We use a multivariate normal jumping distribution centered around $\eta_k^{(t-1)}$ and $\gamma_k^{(t-1)}$, respectively. The cluster assignment update beginning in line 13 is the most time-consuming part of the algorithm - with complexity increasing linearly with sample size. Nevertheless, it is simple to implement with existing statistical software. \\

At each iteration of the sampler, clusters can die (become unoccupied). New clusters can also appear. If the proposed cluster's parameters in line 12 happens to fit a subject's data better than the existing clusters, then the probability in line 15 will dominate those in line 14.\\

We compute $n \times n$ adjacency matrix, $\bm M^{(t)}$, for each posterior draw, $t=1,\dots, T$. The $(i,j)^{th}$ entry, $M_{ij}^{(t)} = I(c_i^{(t)} = c_j^{(t)})$, of this matrix is an indicator for whether subject $i$ was clustered with subject $j$. Taking the element-wise mean of this matrix over Gibbs iterations, $t$, yields an $n \times n$ posterior mode matrix, $\bm M^*$ displaying the frequency with which subject $i$ was clustered with subject $j$. \\

To obtain a hard classification status for each patient, we search for the $\bm M^{(t)}$ that is closest to the posterior mode matrix, $\bm M^*$, in the $L_2$ sense. That is, we search for the posterior $\bm M^{(t)}$ that yield the lowest $\sum_{i,j} (M_{ij}^{(t)} - M^*_{ij}  )^2$. This approach provides an unambiguous way of classifying subjects according to the posterior mode in the presence of label switching. The clusters can then be summarized in terms of their outcome and observed confounder distributions - leading to identification and description of potentially interesting subpopulations of subjects. \\

Moreover, we can view $\bm M^*$ as a posterior adjacency matrix. We could represent this matrix using a network diagram where each subject is represented by a node and the edge between subject $i$ and $j$ has length given by $M^*_{ij}$ - the posterior probability of subject $i$ and $j$ being clustered together.

\section{Simulation Details}

We simulate 1000 data sets under two data generating processes: a clustered setting and parametric setting. For the clustered setting, we simulate 1000 datasets for $i\in 1, \dots, 3000$ patients in the following way:
\begin{itemize}
	\item Draw cluster indicators, $c_i$ with uniform probability $P(c_i=1)=P(c_i=2)=P(c_i=3)=1/3$.
	\item Draw confounder vector $L_i \sim P(L_i | \theta_{c_i}) = N(L_1| \mu_{c_i},\phi_{c_i} ) \prod_{j=2}^5 Ber(L_j | p_{c_i} )$. Where $\theta_{c_i}=(\mu_{c_i},\phi_{c_i}, p_{c_i})$ and the cluster-specific parameters are $\theta_{1}=(.9, .1^2, .25)$, $\theta_{2}=(0, .1^2, .5)$, and $\theta_{3}=(.375, .1^2, .75)$.
	\item Draw treatment $A_i | L_i \sim Ber(p_{c_i}) $, where $p_1 = expit(.5 - 1L_1 + \sum_{j=2}^4 L_j ) $,  $p_2 = expit(.2 + 2L_1 - 2\sum_{j=2}^4 L_j ) $, and  $p_3 = expit(.8 + 1L_1 - \sum_{j=2}^4 L_j ) $.	
	\item Draw structural zero indicator, $Z_i | A_i, L_i \sim Ber(pz_{c_i})$, where  $pz_1 = expit(-2 -2A_i + L_1 + \sum_{j=2}^4 L_j ) $,  $pz_2 = expit(-2 +.5A_i - 2L_1 + 2\sum_{j=2}^4 L_j ) $, and  $pz_3 = expit(2.5 + 1A_i - L_1 - \sum_{j=2}^4 L_j ) $.
	\item If $Z_i=1$, then set $Y_i =0$. Otherwise, draw $Y_i \sim Gamma( shape_{c_i}, scale_{c_i})$ where $scale_1 = 10000$, $scale_2 = 20000$, and $scale_3 = 30000$. The shape parameters are $shape_1 = 4 + A + 2L_1 + \sum_{j=2}^4 L_j $, $shape_2 = 5 + A - 5L_1 + \sum_{j=2}^4 L_j $, and $shape_3 = 7 + A + 1L_1 + \sum_{j=2}^4 L_j $
\end{itemize}

For the parametric setting, 
\begin{itemize}
	\item Draw confounder vector $L_i \sim P(L_i | \theta) = N(L_1| \mu,\phi ) \prod_{j=2}^5 Ber(L_j | p )$. Where $\theta_{c_i}=(\mu,\phi, p) = (0,.1^2, .5 )$.
	\item Draw treatment $A_i | L_i \sim Ber(p) $, where $p = expit(2 + 3L_1 - \sum_{j=2}^4 L_j ) $.	
	\item Draw structural zero indicator, $Z_i | A_i, L_i \sim Ber(pz)$, where  $pz = expit(-2 + .5A_i - 2L_1 + \sum_{j=2}^4 L_j )$.
	\item If $Z_i=1$, then set $Y_i =0$. Otherwise, draw $Y_i \sim Gamma( shape, scale)$ where $scale = \mu/1e9$, $shape = \mu^2/1e9 $, and $\mu = 3e5 + 50000A - 100000L1 + \sum_{j=2}^4 L_j$.
\end{itemize}

We implement BART using the \textbf{BayesTree} package, the doubly robust estimator using the \textbf{twang} package, and code the Gamma hurdle and Gamma +.01 models in \textbf{Stan}. All Bayesian results are based on 5000 posterior draws after 5000 burn-in iteration.

\section{Data Analysis Details}

We take 40,000 posterior draws after allowing for 20,000 burn-in. We initialize the sampler with 5 clusters. Normal priors on continuous covariate distributions. A Normal hyperprior is placed on the mean of this prior with empirical means and standard deviations (scaled by 10 to be slightly wider) . Informative $InvGam(10,10)$ hyperprior placed on the variance. $Beta(1,1)$ priors are placed on binary covariates distributions.\\

A Multivariate normal prior is placed on the outcome regression coefficients. The mean is set to linear regression coefficient estimates using only subjects with positive costs. The prior covariance was set to be diagonal with variance set to the diagonal of the previously mentioned regression's covariance matrix. We scale these variances by 100 to make the prior a little wider. Treatment model and zero-inflation model regresion parameters are also Gaussian, centered around zero with variance of 2. Note that on an odds ratio scale this places most prior mass on regression odds ratios ranging from .014 to 70.\\

Multivariate normal jumping distributions with diagonal variances are used for the Metropolis steps in both the treatment and zero-inflation models. The variance for the jumping distribution was set to .025. Lastly, we choose to estimate the concentration parameter rather than setting it.  We place an $InvGam(1,1)$ prior on the parameter and implement a metropolis step using a Gaussian jumping distribution with variance 1. \\

For standardization, we evaluated the necessary integrals with 30,000 Monte Carlo iterations per posterior draw. The resulting MCMC chains are given

\begin{figure}[H] 
	\centering 
	\includegraphics[scale=.60]{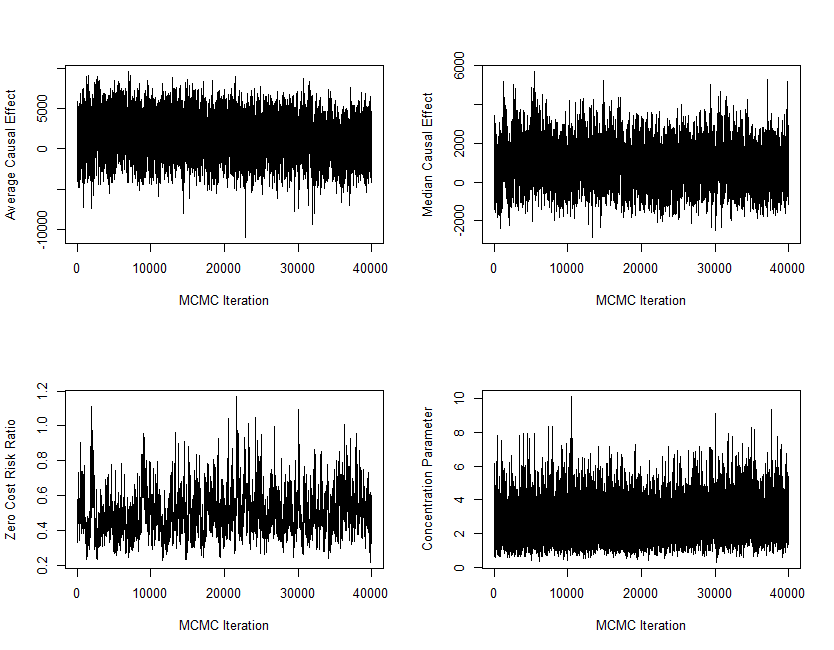}
	\caption{\small 40,000 post-burn-in posterior draws of relevant quantities presented in the data analysis section.}
	\label{fig:patpred}
\end{figure}

\end{document}

%% file: SimSetting_MultiVar.tex

\begin{table}[h!]
	\centering
	\begin{tabular}{lllrrr}
		\hline
		\multicolumn{1}{c}{DGP}     & \multicolumn{1}{c}{Model} &  & \multicolumn{1}{c}{Bias} & \multicolumn{1}{c}{Coverage} & Interval Width \\ \hline
		\multirow{5}{*}{Clustered}  
		& Zero-Inflated DP          &  &      -.081                &            94.3\%         &  21612.2       \\
		& BART                      &  &      -.746                &            76.2\%         &  26374.2       \\
		& Doubly Robust             &  &       .795                &            87.1\%         &  33449.3       \\
		& Gamma Hurdle              &  &      -.509                &            79.8\%         &  19692.2       \\
		& Gamma +.01                &  &      1.817                &             4.7\%         &  27358.1       \\ \hline
		\multirow{5}{*}{Parametric} 
		& Zero-Inflated DP          &  &       .097               &            95.1\%          &    22034.1            \\
		& BART                      &  &      -.054               &            96.1\%          &    23825.3            \\
		& Doubly Robust             &  &      -.027               &            95.9\%          &    23339.1                     \\
		& Gamma Hurdle              &  &      -.014               &            95.1\%          &    21778.7                   \\
		& Gamma +.01                &  &      -.489               &            100\%           &  50580.3                     \\ \hline
	\end{tabular}
	\caption{Results across 1000 simulated datasets. Average bias of the posterior mean is reported as a proportion of the true value ($\Psi= -9740.3$ in the clustered setting and $\Psi=-10184.1$ in the parametric setting). Mean Credible interval widths are presented for the Zero-inflated DP model, the BART model, and two Gamma models. Confidence intervals are given for the doubly robust method.}
	\label{tbl:simres}
\end{table}

%% file: bsl_table.tex
\begin{table}[H]
\centering
\caption{Baseline Characteristics}
\label{tbl:bsl}
\begin{tabular}{lccc} \hline
             & \begin{tabular}[c]{@{}l@{}} Chemotherapy \\ \ \ \  \ \ (n=92)\end{tabular} & \begin{tabular}[c]{@{}l@{}} Radiation Therapy \\ \ \ \ \ \ (n=952)\end{tabular}   & \begin{tabular}[c]{@{}l@{}} SMD \end{tabular} \\ \hline
\multicolumn{1}{l|}{Total Inpatient Costs (\$)}   &                                      22131.59  (28608.07) & 23370.63 (34453.31) &    .039       \\
\multicolumn{1}{l|}{Zero Costs}   &                                      14  (15.2\%) & 75 (7.9\%) &           \\
\multicolumn{1}{l|}{Age (years)}   &                                                      73.68 (6.98)  &      73.25 (5.98)    &     .066                                          \\
\multicolumn{1}{l|}{Household Income (\$)}   &                                   64368.36 (32422.55) & 56785.29 (26166.79)   &     .257                                             \\
\multicolumn{1}{l|}{White}   &                                                   76 (82.6\%)     &     835 (87.8\%)         &     .147                                       \\
\multicolumn{1}{l|}{Diabetic}   &                                                       20 (21.7\%)   &       197 (20.7\%)         &     .026                                       \\
\multicolumn{1}{l|}{CCI}  &                                      &                                      &     .350          \\
\multicolumn{1}{l|}{\ \ \ 0}  &                                                         49 (53.3\%)    &      529 (55.6\%)         &                                             \\
\multicolumn{1}{l|}{\ \ \ 1}  &                                                         22 (23.9\%)    &      260 (27.3\%)         &                                            \\
\multicolumn{1}{l|}{\ \ \ $\geq$2}  &                                                        21 (22.8\%)     &      131 ( 13.8\%)          &                                            \\
\multicolumn{1}{l|}{Grade = 1 } &                                                    	    28 (30.4\%)      &    208 (21.8\%)           &  .196                                        \\
\multicolumn{1}{l|}{FIGO Stage I-N0 or I-A}  &                                                      63 (68.5\%)    &      357 (37.5\%)         &     .653                                    \\ \hline
\multicolumn{4}{p{\wd0+35em}}{\scriptsize Notes: Means and standard deviations are reported for continuous variables. Counts and percentages are reported for categorical variables. All monetary amounts are in 2018 U.S. Dollars.} \\
\end{tabular}
\end{table}

%% file: clust_table.tex
\begin{table}[H]
\centering
\caption{Summary Statistics by Posterior Mode Cluster Assignment}
\label{tbl:clust}
\begin{tabular}{ccccc} \hline
             & \begin{tabular}[c]{@{}c@{}} Red \\ (n=452)\end{tabular} 
             & \begin{tabular}[c]{@{}c@{}} Green \\ (n=160) \end{tabular}   
             & \begin{tabular}[c]{@{}c@{}} Blue  \\ (n=288) \end{tabular} 
             & \begin{tabular}[c]{@{}c@{}} Orange \\  (n=127)\end{tabular} \\ \hline
\multicolumn{1}{l|}{Total Inpatient Costs (\$)}   & 9359.36   & 10560.52 &	23965.20  & 71139.01  \\
\multicolumn{1}{l|}{Radiation}    			      & 91.8\%	& 93.1\%	& 89.9\%		& 89.8\%	\\
\multicolumn{1}{l|}{Age (years)}                  & 72.4	& 73.2	& 73.9	&  74.9	 \\                   
\multicolumn{1}{l|}{Household Income (\$)}        & 74597.17	& 69898.72	& 68377.09	&  64453.8	 \\              
\multicolumn{1}{l|}{White}                        & 90.5\%	& 88.8\%	& 82.6\%	&  85.8\%	 \\                         
\multicolumn{1}{l|}{Diabetic}                     &  0.0\% & 40\% & 38.2\% &  25\% \\        
\multicolumn{1}{l|}{CCI}                          &  & & &  \\        
\multicolumn{1}{l|}{\ \ \ 0}               	      &  100\%    & 15.0\%   & 23.6\%      & 22.0\%  \\
\multicolumn{1}{l|}{\ \ \ 1}                      &  0.0\%    & 85.0\%   & 35.1\%      & 31.5\% \\
\multicolumn{1}{l|}{\ \ \ 2}                      &  0.0\%    & 0.0\%    & 29.5\%      & 15.7\%  \\
\multicolumn{1}{l|}{\ \ \ $\geq3$}                &  0.0\%    & 0.0\%    & 11.7\%      & 30.7\% \\
\multicolumn{1}{l|}{Grade = 1 }                   & 21.7\%    &	25\%     & 21.2\% 	   & 26.8\%  \\                       
\multicolumn{1}{l|}{FIGO Stage I-N0 or I-A}       & 40.3\%	  &39.4\%	 & 38.9\%       & 42.5\% \\          
\hline
\end{tabular}
\end{table}

%% file: DataRes_CausalEsts.tex
\begin{table}[H]
\centering

\begin{tabular}{lccc}
  \hline
                            & Avg. Causal Effect & Median Causal Effect & Causal Risk Ratio of Zero Cost \\ 
  \hline
  Zero-Inflated DP 	        &  \makecell{ 1672.62 \\ (-2566.42, 5722.56)}  & \makecell{872.68  \\ (-833.35, 2790.18)} &  \makecell{ 0.498 \\   (0.31, 0.78) } \\ 
  BART 						&  \makecell{ 1779.62 \\ (-6085.89,  9797.13 )} & -  & - 				 \\ 
  Gamma Hurdle 				&  \makecell{ 2016.71 \\ (-1499.38,  5593.40) } & -  & \makecell{ .505 \\ (.34, .76) } \\ 
  Gamma +.01				&  \makecell{ 4889.00 \\ (1004.37, 8795.61)} & -  & - 				\\ 
   \hline
\end{tabular}
\parbox{6.05in}{\caption{\scriptsize Posterior means and credible interval from standardization procedure using DP mixture along with ATE posterior means and credible intervals from BART, a Gamma hurdle model, and a Gamma model with zeros replaced by +.01 - a common practice among researchers.}} 
\label{tbl:causalests}
\end{table}